\documentclass[10pt, numbers, preprint]{sigplanconf}
\usepackage[utf8]{inputenc}
\usepackage{inconsolata}
\usepackage[T1]{fontenc}

\usepackage{flushend}
\usepackage{amsmath}
\usepackage{enumitem}
\usepackage{graphicx}
\usepackage{hyperref}
\usepackage{listings}
\usepackage{microtype}
\usepackage{ragged2e}
\usepackage{textcomp}
\usepackage{amssymb}
\usepackage{czt}
\usepackage{framed}
\usepackage[usenames,dvipsnames]{color}
\usepackage{setspace}

\let\OldTexttt\texttt
\renewcommand{\texttt}[1]{\OldTexttt{\small #1}}

\definecolor{ZedColor}{cmyk}{1,1,1,0}
\definecolor{ZedBoxColor}{cmyk}{1,1,1,0}

\usepackage[firstpage]{draftwatermark}
\SetWatermarkText{\hspace*{8in}\raisebox{4in}{\includegraphics[scale=0.1]{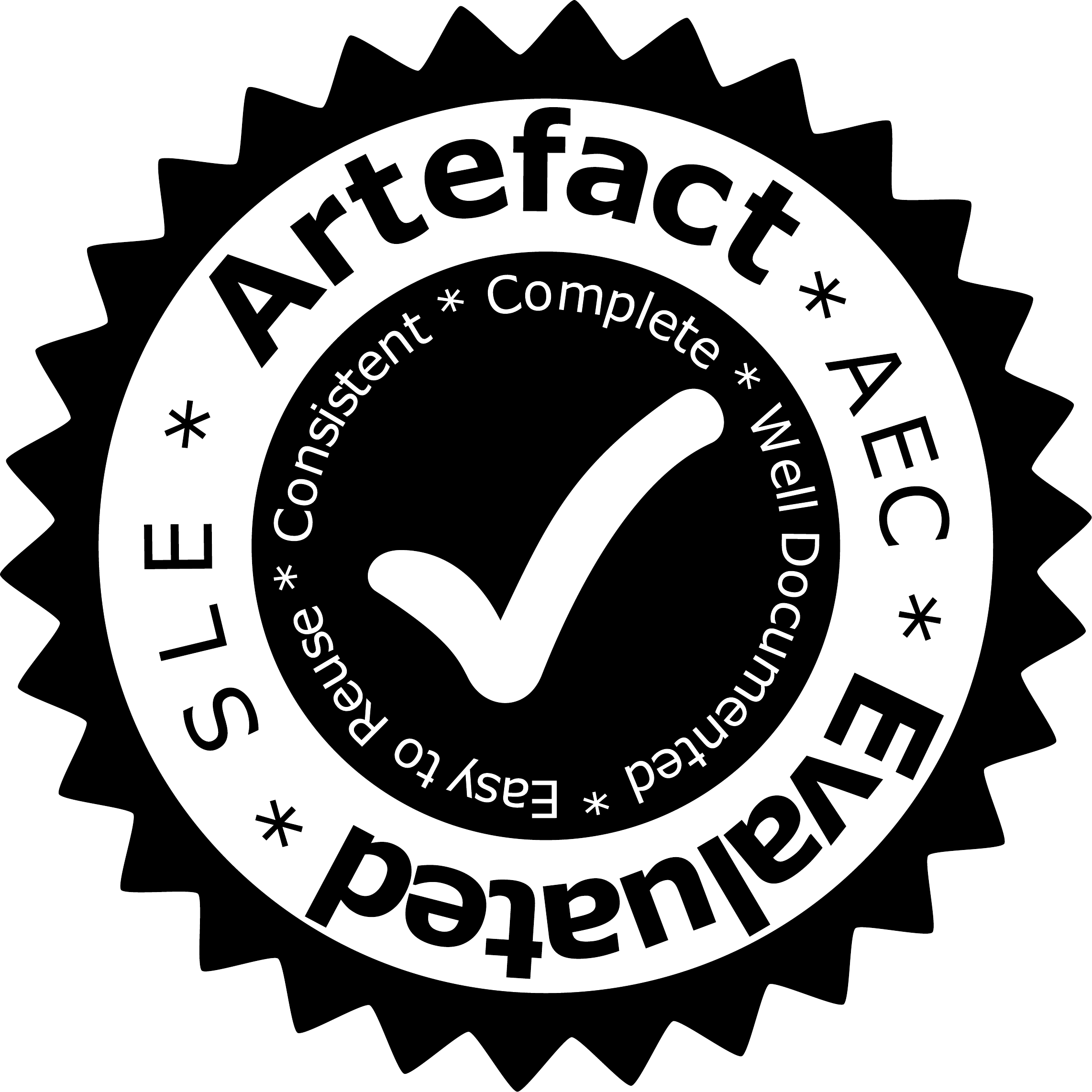}}}
\SetWatermarkAngle{0}

\begin{document}
\toappear{}

\setlength{\pdfpageheight}{\paperheight}
\setlength{\pdfpagewidth}{\paperwidth}

% \conferenceinfo{CONF 'yy}{Month d--d, 20yy, City, ST, Country}
% \copyrightyear{20yy}
% \copyrightdata{978-1-nnnn-nnnn-n/yy/mm}
% \copyrightdoi{nnnnnnn.nnnnnnn}
% \publicationrights{licensed}

\preprintfooter{Taming Context-Sensitive Languages with Principled Stateful
  Parsing}

\title{Taming Context-Sensitive Languages \\ with Principled Stateful Parsing}

\authorinfo{Nicolas Laurent \and Kim Mens}
           {Université catholique de Louvain, ICTEAM}
           {\{nicolas.laurent, kim.mens\}@uclouvain.be}

\maketitle

{\let\thefootnote\relax\footnotetext{Nicolas Laurent is a research fellow of the
    Belgian fund for scientific research (F.R.S.-FNRS).}}

%%%%%%%%%%%%%%%%%%%%%%%%%%%%%%%%%%%%%%%%%%%%%%%%%%%%%%%%%%%%%%%%%%%%%%%%%%%%%%%%

\begin{abstract}
  Historically, true context-sensitive parsing has seldom been applied to
  programming languages, due to its inherent complexity. However, many
  mainstream programming and markup languages (C, Haskell, Python, XML, and
  more) possess context-sensitive features. These features are traditionally
  handled with ad-hoc code (e.g., custom lexers), outside of the scope of
  parsing theory.

  Current grammar formalisms struggle to express context-sensitive features.
  Most solutions lack \emph{context transparency}: they make grammars hard to
  write, maintain and compose by hardwiring context through the entire grammar.
  Instead, we approach context-sensitive parsing through the idea that parsers
  may \emph{recall} previously matched input (or data derived therefrom) in
  order to make parsing decisions. We make use of mutable \emph{parse state} to
  enable this form of recall.

  We introduce \emph{principled stateful parsing} as a new transactional
  discipline that makes state changes transparent to parsing mechanisms such as
  backtracking and memoization. To enforce this discipline, users specify
  parsers using formally specified primitive state manipulation operations.

  Our solution is available as a parsing library named \emph{Autumn}. We
  illustrate our solution by implementing some practical context-sensitive
  grammar features such as significant whitespace handling and namespace
  classification.
\end{abstract}

\category{D.3.4}{Programming Languages}{Processors}[Parsing]

\keywords stateful parsing, grammars, context sensitivity, data
dependence, parsing expressions

%%%%%%%%%%%%%%%%%%%%%%%%%%%%%%%%%%%%%%%%%%%%%%%%%%%%%%%%%%%%%%%%%%%%%%%%%%%%%%%%
\section{Introduction}
\label{intro}

In this section, we review the notion of context-sensitive parsing (\ref{csp}),
describe how we tackle it through recall and stateful parsing (\ref{recall}),
and situate our implementation (\ref{impl}).

In what follows, we use the term \textit{parser} to refer to any unit of
functionality that can match some input text (both at the the lexical and
grammatical levels) and produce a result based on this match. We assume that
simple parsers can be combined into increasingly complex parsers.

\subsection{Context-Sensitive Parsing}
\label{csp}

True context-sensitive parsing is seldom applied to programming languages.
Writing context-sensitive grammars can prove challenging, for instance the
grammar for the language $a^nb^nc^n$ is notoriously
tricky~\cite{parsing-techniques}. In addition, most language features can be
adequately expressed using the much more tractable context-free grammars.
Nevertheless, many mainstream languages exhibit context-sensitive features. Here
are a few examples:

\begin{itemize}

\item In C, in order to determine whether the statement \texttt{x*y;} is the
  product of $x$ by $y$, or rather the declaration of a variable $y$ which is a
  pointer to type $x$, one must analyze the type definitions preceding the
  statement.

\item In Haskell and Standard ML, programmers can introduce operators with
  custom precedence and associativity. The parser needs to interpret these
  definitions in order to be able to parse the remainder of the input.

\item Since Python has significant indentation, a Python parser needs to detect
  when the indentation level increases or decreases.

\item In XML, opening tags must be matched with corresponding closing tags. For
  instance, \texttt{<foo></foo>} is valid while \texttt{<foo></bar>} is not. As
  such, an XML parser must memorize the names of open tags, at arbitrary levels
  of nesting. 

\item Many network protocols, including TCP, make use of length-delimited fields
  whose length is not known in advance but indicated by a length field that
  precedes them. 

\end{itemize}

Most parsing tools cannot adequately handle these syntactic peculiarities,
leading to all sorts of hacks, and sometimes to the rejection of parsing tools
altogether. There are a few exceptions however, which we review in
Section~\ref{related}. Section~\ref{challenges} deals with the key properties
almost all these solutions lack, namely \emph{context transparency}, hence
calling for a new solution.

\subsection{Recall and Stateful Parsing}
\label{recall}

While context sensitivity was first characterized by Chomsky~\cite{chomsky}, his
Context-Sensitive Grammars (CSG) are only of little help, due to the intricate
coding that they require\footnote{The same holds for a large body of work on
  \textit{mildly context-sensitive grammars}.~\cite{mild}}. A CSG is made of
rewrite rules $X \rightarrow Y$ where $X$ and $Y$ are strings of mixed terminals
and non-terminals. These rules must be non-contracting: $Y$, as a string of
symbols, must not be shorter than $X$.\footnote{In reality, CSG rules are not
  required to be non-contracting, but non-contracting grammars and CSG describe
  the same set of languages.~\cite{chomsky63}} As a matter of fact, these
grammars were never meant to describe programming languages, but natural
languages, where the shape of the rules make much more sense. In particular, it
is difficult to encode \textit{recall} constraints: for instance, requiring the
same string of tokens to appear at two different locations in the sentence
(assuming the string is not fixed in advance).

We propose instead to approach context sensitivity through the notion of
\textit{recall}, i.e. the ability to accept sentences based on relationships
between some of their parts. This is more easily understood in parsing terms as
the capability to make parsing decisions based on previously matched input.

We enable recall by allowing users to write parsers which can manipulate mutable
state. However, unlike solutions deployed by existing parsing tools such as
ANTLR~\cite{allstar} and Rats!~\cite{rats}, we are \emph{principled} about state
use. Indeed, parsing algorithms do not proceed linearly. When faced with a
choice, they may speculatively try an alternative, and need to backtrack if this
alternative does not succeed. When backtracking happens, all changes made to the
state during the speculative execution need to be reversed. Parsers may also
memoize the result of a speculative execution. In a stateful model, these
results need to include the state changes incurred by the execution. As will be
explained in Section \ref{stateful-parsing}, we satisfy these requirements by
introducing primitive operations to manipulate mutable state in a principled
way.

\subsection{Implementation}
\label{impl}

We implement our stateful parsing approach as a PEG-like~\cite{peg} (top-down
recursive descent) functional-style~\footnote{Parsers are merely functions.
  Moreover, custom parsers can be defined.} parser-combinator library named
\emph{Autumn}, which can be used with Java, Kotlin, and other Java-compatible
languages. Implementation details are given in Section~\ref{implementation} and
a detailed use case is worked out in Section~\ref{usecase}.

%%%%%%%%%%%%%%%%%%%%%%%%%%%%%%%%%%%%%%%%%%%%%%%%%%%%%%%%%%%%%%%%%%%%%%%%%%%%%%%%
\section{Related Work}
\label{related}

As the problem of context-sensitive features in programming languages is not
new, it is not surprising that several solutions have been proposed. We review
these solutions in order to better put our contributions in perspective. We do
not purport to review the entire body of work on context-sensitive parsing, but
only the approaches closest to our goal. In particular, we left out the
literature on context-sensitive lexical analysis (e.g., \cite{wyk07,
  schrodinger}) which by definition only handles a small subset of all context
sensitivity issues.

%%%%%%%%%%%%%%%%%%%%%%%%%%%%%%%%%%%%%%%%
\subsection{Backtracking Semantic Actions}
\label{backsem}

Parsing with backtracking semantic actions~\cite{backsem} is an approach that
extends a (general) backtracking LR parser with reversible semantic actions.
Upon backtracking, state changes are reversed. Two important restrictions apply:
state changes can only occur during term reduction, and the state can only
affect the parse through semantic conditions that trigger backtracking.

Compared to our approach, the difference in implementation (top-down recursive
versus LR) has far-reaching consequences. Backtracking semantic actions reverse
state changes rather than making snapshots of the state. Accordingly it becomes
impossible to compare state snapshots. These capabilities are very useful in the
context of PEG parsing, as they enable the definition of custom parsers,
left-recursion and memoization; but less so within an LR system where custom
parsers cannot be defined.

We also believe our \textit{top-down recursive-descent} model to be more
intuitive in the presence of state. Logically, state changes occur as parser are
invoked, from left to right and from top to bottom. Backtracking semantic
actions, on the other hand, are executed upon term reduction. This means that a
parser may modify the state before other parsers that matched input on its left.

Despite these caveats, we consider parsing with backtracking semantic
actions~\cite{backsem} to be the safest and most convenient system for
context-sensitive parsing among those presented in this section.

%%%%%%%%%%%%%%%%%%%%%%%%%%%%%%%%%%%%%%%%
\subsection{Data-dependent grammars}
\label{yakker}

Jim et al.~\cite{dependent2010} proposed data-dependent grammars, a formalism
which permits context sensitivity by allowing rules to be parameterized by
semantic values. A parameterized nonterminal appearing on the right-hand side of
a rule acts as a form of function call that also returns a semantic value. These
semantic values are computed by \textit{semantic actions} written in a
general-purpose programming language. There are also \textit{semantic
  predicates} which can make grammar branches succeed or fail depending on a
semantic value.

Data-dependent grammars can be compiled to a format accepted by a target parsing
tool, which must support fairly general semantic actions. In subsequent
work~\cite{dependent2011}, the authors introduced a new kind of automaton that
can be used to implement parsers recognizing data-dependent grammars. These
techniques are put to work in a tool called Yakker.

Data-dependent grammars, though theoretically compelling, suffer from usability
issues. The value-passing model means that the parse state needs to be threaded
throughout the grammar. Making a rule dependent on a new semantic value means
that all rules through which this rule is reachable might need to be modified to
pass this value around. Maintainability-wise, this is far from ideal. Moreover,
it harms composability, as a rule must be aware of all states it has to pass
through.

Afroozeh and Izmaylova~\cite{dependent-transform} show how advanced parser
features such as lexical disambiguation filters, operator precedence,
significant indentation and conditional preprocessor directives can be
translated to data-dependent grammars. Quite clearly, the task is non-trivial
and one comes away with the feeling that dependent grammars are better suited as
an elegant calculus to be targeted by parsing tool writers rather than as a
paradigm that fits the needs of tool users. The machinery implementing the
formalism is also distinctively non-trivial, involving a multi-stage
transformation into a continuation routine or into a new kind of automaton. In
contrast, our approach consists of a lightweight library that can be layered on
top of a general-purpose programming language.

Finally, we note that the much older Definite Clause Grammars (DCGs)~\cite{dcg}
formalism works on almost exactly the same principle, but building upon logic
programming. Accordingly, it suffers from similar limitations.

%%%%%%%%%%%%%%%%%%%%%%%%%%%%%%%%%%%%%%%%
\subsection{Monadic parsers}
\label{monadic}

Monadic parsing~\cite{monadic} is a well-known way to build functional-style
parser-combinator libraries, made popular by Haskell libraries such as
Parsec~\cite{parsec}. In this paradigm, the type of a parser is a function
parameterized by a result type, i.e. with signature
$string \rightarrow (string, result)$, where the parameter string is the input
text and the returned string is the input remaining after parsing. The parser
type is also a monad instance, meaning there is a \texttt{bind} function whose
signature, in Haskell notation, is:

\texttt{Parser r1 -> (r1 -> Parser r2) -> Parser r2}\\
where \texttt{r1} and \texttt{r2} are result types. This function takes a parser
as first parameter, and a function which transforms the result of the parse into
another parser as second parameter. When invoked, the parser returned by
\texttt{bind} will invoke the first parser, pass its result (of type
\texttt{r1}) to the function, then invoke the parser this function returns,
yielding a result of type \texttt{r2}. 

The important point about monadic parsers is that they can handle context
sensitivity. Indeed, the second parameter to \texttt{bind} (the function)
returns a parser from a result. This means that the behaviour of the parser
returned by \texttt{bind} depends on data acquired during the parse: this is a
form of \textit{recall}.

An in-depth analysis of this aspect was done by Atkey~\cite{semsem}. In
particular, he formalizes monadic parsers by introducing \textit{active
  right-hand sides}, which are the right-hand sides of rules that can contain
monadic combinators. These combinators generate grammar fragments at parse-time
(much like a monadic parser generates a new parser), hence the term
\textit{active}. While monadic parsing seems at first sight very similar to the
data-dependent grammars from Section~\ref{yakker}, Atkey~\cite{semsem} carefully
contrasts the two approaches:

\begin{quotation}\noindent\emph{ We characterise their [Jim et al.]
    approach as refining context-free grammars: each Yakker grammar has an
    underlying context-free grammar with regular right-hand sides, and the
    constraints allow for sophisticated data-dependent filtering of parses. In
    contrast, we consider active right-hand sides that generate the grammar as
    the input is read.}
\end{quotation}

Nevertheless, monadic parsers suffer from the same pitfalls as data-dependent
grammars: the state is threaded through the grammar (or code), leading to poor
maintainability and composability.

%%%%%%%%%%%%%%%%%%%%%%%%%%%%%%%%%%%%%%%%
\subsection{Attribute Grammars}

Attribute grammars~\cite{attributes} associate attributes to AST nodes (assuming
an AST node per matched grammar rule). The attributes can be synthesized: their
value derived from the attributes of subnodes, or inherited: their value
computed by a parent node. The formalism supports context-sensitive parsing
through production guards predicated over attributes.

However, attribute grammars are not context-transparent. To enable recall, they
need to propagate the recalled value from the definition site to the use site,
through a chain of of synthesized and inherited attributes. Even reference
attributed grammars~\cite{rag}, which allow attributes to contain references to
nodes, do not fully solve this distribution problem.

%%%%%%%%%%%%%%%%%%%%%%%%%%%%%%%%%%%%%%%%
\subsection{Stateful Parsing}
\label{stateful}

Manipulating parse-wide state can be an effective solution to the problem of
data dependence: the data depended upon can be written in the state when
encountered and read or even altered later on.

Broadly speaking, we can distinguish two big classes of stateful parsing tools.
First, there are parser combinator libraries that allow users to write their own
subparsers. Notable examples include Parboiled~\cite{parboiled}, Lua
Peg~\cite{luapeg} and Scala's parser combinators~\cite{scala-combinators}. Since
these custom parsers are implemented in a general-purpose programming language,
they can manipulate state, even though the libraries make no provision for this.
Second, there are parsing tools that provide very general semantic actions and
semantic predicates. Notable examples include Bison~\cite{bison} and
ANTLR~\cite{allstar}. These work much like their counterpart in Yakker (cf.
Section~\ref{yakker}), except that instead of returning a value, semantic
actions may modify a global state object.

Unfortunately, most parsing tools in both categories do not make the necessary
provisions for dealing with backtracking and memoization: if the parser
backtracks over a construct that made state changes (semantic action or custom
parser), these changes need to be undone; if the parser can memoize the result
of a construct, state changes need to be memoized as well. In the absence of
such guarantees, a construct can only access state which it is sure has not been
corrupted by changes that should have been discarded. It must also be sure that
some state-altering construct was not skipped due to memoization. These are
tricky propositions to verify even for medium-sized grammars, and every change
to the grammar threatens to falsify them.

One may think that solving the backtracking problem is simply a matter of
inserting a construct that reverses state changes whenever a rule fails.
However, a rule can be backtracked over even if it succeeded. It suffices that
one of the rules through which our rule was reached fails. Hence this scheme
would entail, for each state-altering construct, the modification of every rule
through which it can be reached.

%%%%%%%%%%%%%%%%%%%%%%%%%%%%%%%%%%%%%%%%
\subsection{Rats!}

Rats!~\cite{rats} is a fully-memoizing (\textit{packrat}) PEG parser. Rats! is,
to the best of our knowledge, the only stateful parsing tool that provides some
guarantees for state usage, by ensuring that state changes are discarded if
certain conditions are met.

For this purpose, Rats! introduces \textit{transactions} that wrap rules under
which state changes might occur. A transaction can either succeed, in which case
its state changes are retained, or fail, in which case the changes are
discarded. Rats! also requires that a nonterminal invoked at a given position
within a transaction must always modify the state in the same way, no matter how
that nonterminal was reached. Combining transactions with this requirement
ensures that Rats! will never have to discard the memoization of a rule, hence
upholding the linear-time guarantee of packrat parsers.

In spite of its advantages, this scheme has two important pitfalls. First, it
requires nonterminal invocations at a given position to always return the same
result. This precludes parsing expressions that modify the behaviour of the
parsing expression they invoke. However, this capability is valuable in
practice. For example, we use it to enable left-recursion handling in our
library (cf. Section~\ref{leftrec}).

Second, state changes are not memoized. If a rule succeeds after applying a
state change, but the enclosing transaction fails, the changes are lost. If we
wanted to call the rule at the same position again, the memoized result would be
used and it does not include the state changes. This means that a state change
cannot safely be referenced by two different transactions, and that transactions
cannot be re-tried after a state change higher up in the grammar hierarchy.

%%%%%%%%%%%%%%%%%%%%%%%%%%%%%%%%%%%%%%%%
\section{Context Transparency}
\label{challenges}

As the previous section has shown, enabling the definition of context-sensitive
languages without jeopardizing maintainability, composability or even safety is
no easy feat. We put forward the notion of \emph{context transparency} as the
gold standard that a context sensitive parsing mechanism needs to meet in order
to be considered sufficiently practical.

\begin{framed}
  A grammatical construct is \textbf{context-transparent} if it is unaware of
  the context shared between its ancestors and its descendants.
\end{framed}

Data-dependent grammars, monadic parsers, DCGs and attribute grammars are not
context-transparent because of the need to explicitly pass values around. For
instance, consider two data-dependent grammars\footnote{The same reasoning
  applies to monadic parsers, DCGs and attribute grammars.}: a grammar for a
Python-like language with significant indentation, in which the rules for
block-level constructs (statements, definitions) are paremeterized by the
indentation level; and a grammar for a generic macro definition language (e.g.,
GNU M4). We want to compose these two languages such that macro definitions may
appear anywhere where definitions can appear in our Python-like language.
Additionally, we want macro bodies to include Python-like code.

The issue is that the rules in the macro language grammar know nothing about
indentation level, yet the indentation level needs to be shared between the
block holding the macro definition and the Python-like code appearing inside
macro definitions. In this case, the lack of context transparency would force us
to rewrite all rules in the macro language grammar to carry around the
indentation level.

Stateful parsers also are not context-transparent, as they must ensure that no
unforeseen backtracking or memoization takes place. For instance, if a parser
$a$ manipulates the state and its callers do not expect it to backtrack, it
cannot be swapped for a parser $c(a)$ (where $c$ is some parser combinator)
without first ensuring that $c(a)$ never backtracks over $a$.

Lack of context-transparency makes grammars hard to reason about, hence hard to
write and to maintain: refactoring, extending or composing grammars becomes
particularly challenging, because each change to a rule might entail the need to
modify all rules through which it is (transitively) reachable. In stateful
parsers, such changes are liable to introduce undesired backtracking or
memoization.

We suggest a simple solution: use stateful parsing (which does not thread
context through the grammar), but undo state changes upon backtracking and allow
the memoization of state changes. And to achieve this, we introduce a new
context sensitivity handling discipline: \emph{principled} stateful parsing.

%%%%%%%%%%%%%%%%%%%%%%%%%%%%%%%%%%%%%%%%%%%%%%%%%%%%%%%%%%%%%%%%%%%%%%%%%%%%%%%%
\section{Principled Stateful Parsing}
\label{stateful-parsing}

In Section~\ref{intro}, we established the relevance of context-sensitive
parsing and introduced the notion of \textit{recall} as a way to express
context-sensitive features in terms of backreferences to previously matched
input. We enable recall by storing the matched input (or data derived thereof)
in a mutable data store: the \textit{parse state}. This section expounds how
\emph{principled stateful parsing} is able to work with parse state while
avoiding the usual pitfalls of stateful parsing (cf. sections~\ref{stateful} and
\ref{challenges}).

%%%%%%%%%%%%%%%%%%%%%%%%%%%%%%%%%%%%%%%%
\subsection{Intuition}
\label{intuition}

Before diving into a formal explanation, we present the remarkably simple
intuition behind the approach.

The point of using state is to pass context around implicitly, without the need
to hardwire context in the grammar, hence achieving context transparency (cf.
Section \ref{challenges}).

If the execution of a parser were linear, simply reading/writing to this state
would suffice. Unfortunately, parsers must sometimes perform speculative
executions that may fail further down the line, a phenomenon called
backtracking. When backtracking occurs, all state changes in the speculative
execution being backtracked over must be reversed. Hence, we need an operation
that can take a \textbf{snapshot} of the state at a given point, and an
operation that can \textbf{restore} the state described by such a snapshot.

Given these requirements, it helps to think of the parse state as a log of the
operations applied to the state, which can be snapshot and rolled back as
required. Appropriately, this is also how we formalize the parse state.

Additionally, it is sometimes desirable to save the result of a speculative
execution (whether it failed or not), i.e., the state changes it induced: a
\textit{delta} acquired by performing a \textbf{diff} between the states before
and after the execution. It is also necessary to be able to \textbf{merge} these
changes back into the state. The most straightforward application of the
\textit{diff} and \textit{merge} capabilities is the memoization of parse
results. However, other valuable use cases exist, such as longest-match parsing
and left-recursive parsing (see Section~\ref{leftrec}).

This motivates the need for four primitive state-manipulation operations:
\textbf{snapshot}, \textbf{restore}, \textbf{diff} and \textbf{merge}. These
operations are described in section \ref{operations}.

\begin{framed}
  \textbf{Principled stateful parsing} is an approach where parsers behave
  transactionally: each parser invocation either succeeds or leaves the state
  untouched. Additionally, it is possible to generate and merge deltas
  corresponding to state changes made by parser invocations. All this is made
  possible through the use of formally specified state manipulation operations.
\end{framed}

%%%%%%%%%%%%%%%%%%%%%%%%%%%%%%%%%%%%%%%%
\subsection{Formalization}
\label{formal}

We formalize our approach using the Z notation~\cite{zed}, though eschewing its
schema calculus in favor of a purely functional presentation.\footnote{To
  improve the presentation, we took some liberty with the Z layout (but not with
  the notation). A machine-understandable version of the specification is
  available online~\cite{sle2016artif}.} The Z notation is a formal
specification language that builds on top of Zermelo-Frankel set theory,
first-order logic and simply typed lambda calculus. As such, Z can be seen as a
language where functions can be defined in lambda calculus extended with
predicates from first-order logic and set theory. Formal assertions over the
functions can be made using the same notation. We also note that in Z, all types
used in the lambda calculus are sets.

Since we adopt the functional parser-combinator approach (cf.
Section~\ref{intro}), parsers are simply functions manipulating parse state
(Section~\ref{pstate}) whose set-theoretic signature is given in
Section~\ref{parsers}. Section~\ref{operations} formally specifies the primitive
state-manipulation operations that were briefly introduced in Section
\ref{intuition}. Finally, Section~\ref{call} gives the semantics of parser
invocation by specifying the $call$ operation, which maps a parser (as defined
in \ref{parsers}) to a single state transformation.

%%%%%%%%%%%%%%%%%%%%%%%%%%%%%%%%%%%%%%%%
\subsubsection{Parse State}
\label{pstate}

At the core of our approach lies the notion of parse state. The parse state
abstracts over a general mutable data store. We do not place any constraint on
the data within the store. This is formalized as follows.

\vspace{-0.4cm}
\begin{axdef}
[ CHANGE ]\\
STATE = \seq CHANGE
\end{axdef}

The square brackets introduce the abstract set $CHANGE$ of all state changes.
What exactly constitutes a state change (most likely the mutation of a memory
location) is an implementation concern that is not relevant to the
formalization.

$STATE$ is the set of possible parse states: i.e., of possible configurations of
our mutable store. We represent a parse state as a sequence of state changes.
This means that a state can be seen as a log of the operations over the mutable
store it represents, assuming some well-defined initial state.

In Z, the set of sequences of items from the set $S$ is written $seq ~~ S$ and
corresponds to the powerset of pairs $(i, s) \in \nat \times S$, or equivalently
to the powerset of partial functions $\nat \pinj S$. In each sequence, the
indices are unique and consecutive.

In practice, an implementation of the approach will want to use parse state to
reify important parsing notions, such as input position. We consciously avoided
making our formalism needlessly specific, hence the absence of some usual
parsing notions such as input position. This enables using our approach to parse
non-linear inputs (e.g., object graphs), or perform computations that only bear
nominal resemblance to traditional parsing, even though this direction is
outside the scope of the current paper.

%%%%%%%%%%%%%%%%%%%%%%%%%%%%%%%%%%%%%%%%
\subsubsection{Parsers}
\label{parsers}

A parser represents a computation over the parse state that either succeeds or
fails, and has side effects on the parse state, in the form of \textit{state
  changes}, as introduced in the previous section.

\vspace{-0.4cm}
\begin{axdef}
TRANSFORM = STATE \fun STATE \\
PARSER = STATE \fun \seq TRANSFORM \\
RESULT ::= success | failure \\
result : STATE \fun PARSER \fun RESULT
\end{axdef}

Formally, a parser is a function from a state --- the current state at the time
of invocation --- to a sequence of transformations, which move from one state to
another. This amounts to defining a parser in terms of its execution trace.

Two things seem to be missing from this definition. First, it does not say if
the parse succeeds when run over a specific state. This property is exposed
separately through the $result$ predicate rather than as part of the $PARSER$
signature. This approach is not significant: it simply makes the math look
nicer. Second, the input being parsed does not explicitly appear in the
signature. Instead, the input is assumed to be held within the parse
state.\footnote{Nothing precludes the input from being mutable, even though we
  have not investigated the usefulness of the idea.}

% TAG SHORTEN

A parser is a recognizer of states. It accepts states for which
$result ~ state = success$ holds. If within the input state one dissociates the
\textit{parse input} from the rest of the state (the \textit{context}), one can
see that the parser recognizes --- hence also defines --- different languages
depending on the context.

But a parser is also a transformer of states as well: when invoked it performs a
$STATE \rightarrow STATE$ transformation. In section \ref{call} we explain how
to derive this transformation from a parser (recall that parsers have type
$PARSER$ defined as $STATE \rightarrow seq\ TRANSFORM$), as a means of defining
the semantics of a parser given its execution trace. We could alternatively have
defined $PARSER$ as $STATE \rightarrow TRANSFORM$ (with the result being the
composition of the transformations in the sequence), or directly as
$STATE \rightarrow STATE$. We chose to emphasize the execution trace --- a
sequence of transformations --- instead, because the primitive state operations
described in the next section are suppliers of such transformations, to be
composed to yield the transformation performed by the parser.

This representation also emphasizes that the parse state is both an input of the
parser and an input of the returned transformations. This reflects the fact that
a parser is context-sensitive: it chooses which operation to perform depending
on the state. This is closely related to the notions of active right-hand sides
\cite{semsem} and monadic parsing \cite{monadic}. In fact, each operation in the
sequence is chosen depending on the state obtained by running the initial state
through the composition of all preceding transformations. Abstracting over this
makes the specification much simpler, without altering its meaning.

%%%%%%%%%%%%%%%%%%%%%%%%%%%%%%%%%%%%%%%%
\subsubsection{Primitive Operations}
\label{operations}

We now present six primitive operations (amongst which the four announced in
Section~\ref{intuition}) that parsers can perform.

\vspace{-0.4cm}
\begin{axdef}
SNAPSHOT = \seq CHANGE \\
DELTA = \seq CHANGE \\
\\
call : PARSER \fun TRANSFORM \\
snapshot : STATE \fun STATE \\
diff : SNAPSHOT \fun STATE \fun DELTA \\
applyChange : CHANGE \fun TRANSFORM \\
restore : SNAPSHOT \fun TRANSFORM \\
merge : DELTA \fun TRANSFORM
\end{axdef}

\paragraph{Call} Of these six, $call$ has a special status: it represents the
invocation of a parser. We will define this operation in section \ref{call},
hence specifying the semantics of parsers given their execution trace. Note that
the signature definition of $call$ expands to
$PARSER \rightarrow STATE \rightarrow STATE$: a parser must be called with a
state as parameter.

\paragraph{Snapshot} A snapshot, as the name implies, is a capture of the state
at a specific point during the execution. Naturally, this makes $SNAPSHOT$, the
set of all snapshots, equivalent to $STATE$. Formally, the $snapshot$ operation,
which creates such a capture, is simply the identity function.

\vspace{-0.4cm}
\begin{axdef}
snapshot = \lambda x : STATE @ x
\end{axdef}

\paragraph{Diff} The $diff$ operation returns a $DELTA$ object representing the
difference between a snapshot and the current state, as a set of state changes.
As a precondition, this operation requires the snapshot it receives to be a
prefix of the current state. This is expressed with the Z built-in $prefix$
infix operator. By keeping the deltas append-only, we ensure that a delta can be
later \textit{merged} to any state, not just the one corresponding to the
snapshot.

\vspace{-0.4cm}
\begin{axdef}
\forall sn : \dom diff @ \forall st : \dom ~ ( diff~sn ) @ \\
    \hspace{15pt} sn~prefix~st
\end{axdef}

Since deltas are state suffixes, $DELTA$, the state of all deltas, is equivalent
to $STATE$.

Assuming the precondition is respected, $diff$ can be defined as the remainder
of the current state after chopping off the prefix corresponding to the
snapshot. In Z, the $squash$ function packs the indices (left-hand side) of a
set of pairs in $\nat \times S$, where $S$ is some set, in order to turn this
set into a proper sequence. For instance, it turns $\{ (2, x), (5, y) \}$ to
$\{ (1, x), (2, y) \}$.

\vspace{-0.4cm}
\begin{axdef}
diff = \lambda sn : SNAPSHOT @ \lambda st : STATE @ \\
    \hspace{15pt} squash ~ ( st \setminus sn )
\end{axdef}

\paragraph{Transformations} All operations except $diff$ and $snapshot$ return a
transformation. Recall that we defined $PARSER$ as
$STATE \rightarrow seq ~ TRANSFORM$. The transformations returned by the
operations are precisely those which will be part of a parser's execution trace.
$diff$ and $snapshot$ are different because they do not modify the parse state.
Instead, $diff$ and $snapshot$ create new objects, which can be freely passed
through the parse state.

\paragraph{ApplyChange} The $applyChange$ operation is very simple: given a
change, it simply returns a transformation that applies this change, by
appending it to the change log. It can be defined as follows, using the
concatenation operator ($\smallfrown$) to append the change to the old log.

\vspace{-0.4cm}
\begin{axdef}
applyChange = \lambda c : CHANGE @ \lambda st : STATE @ \\
    \hspace{15pt} st \cat \langle c \rangle
\end{axdef}

\noindent This ``operation'' models the fact that parsers can perform arbitrary
state changes.

\paragraph{Restore} The $restore$ operation takes a snapshot as input and
returns a transformation that brings the state to that described by the
snapshot.

\vspace{-0.4cm}
\begin{axdef}
restore = \lambda sn : SNAPSHOT @ \lambda st : STATE @ sn
\end{axdef}

\paragraph{Merge} The $merge$ operation takes a delta as input and returns a
transformation that appends this delta to the input state.

\vspace{-0.4cm}
\begin{axdef}
merge = \lambda d : DELTA @ \lambda st : STATE @ st \cat d
\end{axdef}

%%%%%%%%%%%%%%%%%%%%%%%%%%%%%%%%%%%%%%%%
\subsubsection{Parser Invocation Semantics}
\label{call}

We now look at how the transformation returned by the $call$ operation can be
derived from the execution trace returned by a parser. Recall that the $call$
operation's signature is $PARSER \rightarrow TRANSFORM$.

We start by defining two helper functions. $composeTwo$ maps sequences of
transformations of length $n \geq 2$ to a sequence of length $n-1$ similar to
the input sequence, but where the first two items have been replaced by their
composition ($s~1$ and $s~2$ access the first two items of $s$ while $\comp$ is
the relational composition operator). $reduceN$ takes a natural $n$ and a
sequence of transformations and returns the composition of its $n$ first items,
or the identity transformation if $n = 0$. This is achieved by iteratively
running the sequence through $composeTwo$, using the Z built-in $iter$ operator.

\vspace{-0.4cm}
\begin{axdef}
composeTwo = \lambda s : \seq TRANSFORM @ \\
    \hspace{15pt} \langle s ~ 1 \comp s ~ 2 \rangle \cat tail ~ ( tail~s ) \\
reduceN = \lambda n : \nat @ \lambda s : \seq TRANSFORM @ \\
    \hspace{15pt} \IF ~ ( n = 0 ) ~ \THEN ~ \id STATE\\
    \hspace{15pt} \ELSE ~ iter ~ ( n - 1 ) ~ composeTwo~s~1 \\
\end{axdef}

With this in place, we define the result of $call$ as the composition of all
transformations within the call's execution trace, assuming the parser
invocation is successful. Otherwise, the identity transformation is returned.
The hash sign ($\#$) is an operator returning the cardinality of a set.

\vspace{-0.4cm}
\begin{axdef}
call = \lambda p : PARSER @ \lambda st : STATE @ \\
    \hspace{15pt} \IF ~ ( result~st~p = success ) \\
        \hspace{30pt} \THEN ~ reduceN ~ ( \# p~st ) ~ ( p~st ) ~ st \\
    \hspace{15pt} \ELSE ~ st
\end{axdef}

%%%%%%%%%%%%%%%%%%%%%%%%%%%%%%%%%%%%%%%%%%%%%%%%%%%%%%%%%%%%%%%%%%%%%%%%%%%%%%%%
\section{Implementation}
\label{implementation}

We implemented the \emph{principled stateful parsing} approach in a
general-purpose parsing library called \textit{Autumn}. It is freely available
online~\cite{sle2016artif}. Autumn is implemented in Kotlin, an up-and-coming
JVM language that closely matches Java's semantics while reducing boilerplate.
Kotlin possesses many features that make it particularly well suited for writing
domain-specific languages (DSLs), an ability we exploit to define grammars. We
will introduce these features as we encounter them. Our approach is not
language-specific and can easily be ported to other languages.

We start by exposing the fundamentals of the Autumn API and how it relates to
our formalization (Section~\ref{autumn-api}). We then show the API in action on
a simple example (Section~\ref{parser-example}). Finally we discuss how the API
enables simple left-recursion handling (Section~\ref{leftrec}).

%%%%%%%%%%%%%%%%%%%%%%%%%%%%%%%%%%%%%%%%
\subsection{The Autumn API}
\label{autumn-api}

In this section, we review how our implementation relates to our formalization
of principled stateful parsing (Section \ref{stateful-parsing}). Figure
\ref{key-types} shows the key interfaces and classes in our implementation.

\begin{figure}[!ht]
\begin{lstlisting}
interface Parser {
    fun parse (ctx: Context): Result
}

sealed class Result {
  object Success: Result()
  open class Failure (val pos: Int, val msg: String)
    : Result()
}

class Context (input: String,
               vararg states: State<*, *>) {
  var pos: Int = 0
  val text: String = input + '\u0000'
  fun <T: State<*,*>> state(klass: Class<T>): T { ... }

  fun snapshot(): Snapshot { ... }
  fun restore(snap: Snapshot) { ... }
  fun diff(snap: Snapshot) { ... }
  fun merge(delta: Delta) { ... }

  ...
}

class Snapshot { ... }
class Delta { ... }

interface State<Snapshot, Delta> {
  fun snapshot(): Snapshot
  fun restore(snap: Snapshot)
  fun diff(snap: Snapshot): Delta
  fun merge(delta: Delta)
}

abstract class Grammar {
  open val whitespace: Parser
    = ZeroMore(CharPred(Char::isWhitespace))
  open val root: Parser
  open val requiredStates: List<State<*,*> = emptyList()
  ...
}
\end{lstlisting}
\caption{Key interfaces and classes in the implementation of Autumn.}
\label{key-types}
\end{figure}

\vspace{0.3cm} \noindent \textbf{Parser} We represent a parser by an instance of
the \texttt{Parser} interface. Implementers must override the \texttt{parse}
method, which takes a \texttt{Context} as parameter and returns a
\texttt{Result}: either a \texttt{Success} or a \texttt{Failure} which holds the
position at which the failure occurred, together with a diagnostic message. It
can also hold custom diagnostic information through subclassing of
\texttt{Failure}. This gives a lot of control over the error messages that will
be shown to the user.

\vspace{0.3cm} \noindent \textbf{Context} Each parse --- the invocation of a
parser on a complete piece of input text --- has an associated \texttt{Context}
object. The role of this object is to hold the state for the parse. The context
is passed down to parsers during parser invocation, so that all parsers may
access it. Using a context object rather than global state allows multiple
parses to co-exist, potentially in parallel.

The mutable store mentioned in the formalization is represented as a collection
of singleton classes implementing the \texttt{State} interface. Parsers can
retrieve a state instance by calling the \texttt{state} method with the proper
class object. Note that the state held by the context also includes the input
text and the input position, although, as a special provision, these can be
accessed directly through the \texttt{text} and \texttt{pos} properties
respectively.

\vspace{0.3cm} \noindent \textbf{State} The \texttt{State} interface has four
methods: \texttt{snapshot}, \texttt{restore}, \texttt{diff} and \texttt{merge},
corresponding to the four key operations introduced in Section \ref{intuition},
but only locally for the \texttt{State} instance itself. To get the parse-wide
semantics of Section \ref{stateful-parsing}, we aggregate the state operations
over all \texttt{State} instances. This is achieved through the methods in the
\texttt{Context} class that mimic the \texttt{State} interface. These methods
manipulate the \texttt{Snapshot} and \texttt{Delta} classes (not to be confused
with the eponymous type parameters of interface \texttt{State}), which aggregate
\texttt{State}-level snapshots and deltas.

In our formalization, we represented the mutable store as a log of all
operations over the store. In practice, this might not be a good idea, as
parsers must compute using the state, meaning the ``current state'' would need
to be re-derived from the whole log at least every time backtracking happens,
unless all operations were fully reversible.

The approach we took instead was to give maximum implementation flexibility to
the programmer: he can choose, for each \texttt{State} instance, the most
appropriate strategy to create snapshots and deltas, as well as to restore/merge
them back in. Maintaining a log of reversible operations is only one possibility
among many.

However, having the programmer implement the \texttt{State} interface for each
data structure he wishes to manipulate would be tedious and repetitive. Hence,
we supply base implementations for common use cases, such as:

\begin{itemize}

\item \texttt{CopyState}: for states that are records containing just a few
  fields, which we can afford to copy every time, and which can be treated as a
  unit (i.e., a delta cannot represent that a field changed while the others
  retained their previous value --- the value of every field is systematically
  captured).

\item \texttt{StackState}: represents a stack as a singly linked list (which is
  naturally immutable). Snapshots and deltas are represented as nodes in the
  list. The list is treated as a unit.

\item \texttt{MonotonicStack}: Similar to \texttt{StackState}, but adds the
  restriction that \texttt{diff} must only be called with a snapshot that is a
  suffix of the current stack. Deltas are then prefixes of the stack and can be
  grafted back at a later time, allowing for granular change handling.

\item \texttt{MapState}: represents a map as an immutable Hash Array Map Trie
  (HAMT)~\cite{hamt}. Our implementation is based on that of Steindorfer and
  Vinju~\cite{hamt-fast} which ensures good performance. The map is treated as a
  unit.

\item \texttt{InertState}: represents state that does not change during the
  parse, or whose change is not significant (e.g., logging logic). All
  operations are implemented as no-ops.

\end{itemize}

\noindent\textbf{Grammar} Programmers must subclass the \texttt{Grammar}
class in order to define a new grammar. When doing so, they must define the root
parser by overriding \texttt{root()}, and provide any required \texttt{State}
instances by overriding \texttt{requiredStates()}. \texttt{Grammar} also defines
a default \texttt{whitespace} parser which consumes any number of characters
matching the Java \texttt{Char::isWhitespace} predicate.

Parser combinator libraries traditionally struggle with the definition of
recursive parsers: it is forbidden to write [\texttt{val A = Seq(..., A)}]
because \texttt{A} is not defined yet when it is evaluated on the right-hand
side. Using the \texttt{Grammar} initialization logic, we can replace the
recursive reference to \texttt{A} with the [\texttt{!"A"}] operator-overloading
syntax: [\texttt{val A = Seq(..., !"A")}]. This creates a stub parser which will
be patched with a reference to \texttt{A} at parse-time. To achieve this, the
\texttt{Grammar} class maps names to parsers through reflection over its
\texttt{Parser}-valued fields.

%%%%%%%%%%%%%%%%%%%%%%%%%%%%%%%%%%%%%%%%
\subsection{Contract}

Autumn enforces the \textit{principled stateful parsing} guarantees, but only if
its interfaces are implemented correctly. In particular:

\begin{itemize}

\item Each implementation of \texttt{Parser} must either succeed or undo all the
  state changes it incurred. This is usually achieved through the use of
  \texttt{snapshot()} and \texttt{restore()}.

\item Each implementation of \texttt{State} must implement its operations
  according to the specification given in Section \ref{stateful-parsing}.

\end{itemize}

%%%%%%%%%%%%%%%%%%%%%%%%%%%%%%%%%%%%%%%%
\subsection{Example}
\label{parser-example}

As an illustration, Figure~\ref{seq-parser} shows the implementation of one of
the most fundamental parser combinators, the sequential composition of parsers.
The resulting parser calls its subparsers sequentially, succeeding if they all
succeed. If one of them fails, the parser reverts the state to its initial
condition.

\begin{figure}
\begin{lstlisting}
class Seq (vararg children: Parser): Parser(*children) {
  override fun _parse_(ctx: Context): Result {
    val snapshot = ctx.snapshot()
    for (child in children) {
      val r = child.parse(ctx)
      if (r is Failure) {
        ctx.restore(snapshot)
        return ctx.failure
    }   }
    return Success
}   }
\end{lstlisting}
  \caption{Implementation of the sequential parser combinator in Autumn.}
\label{seq-parser}
\end{figure}

%%%%%%%%%%%%%%%%%%%%%%%%%%%%%%%%%%%%%%%%
\subsection{Left-Recursion}
\label{leftrec}

When a parser invokes itself (either directly or indirectly through intermediate
parsers) without intervening state changes, the result is an infinite loop of
parser invocations. This is a well-known problem of top-down recursive parsers,
called \textit{left-recursion}. Fortunately, it can be mitigated as follows:
start by running the left-recursive parser while failing all left-recursive
invocations, then re-run it, using the result of the initial parse as the result
of all left-recursive invocations. Repeat until as much input as possible has
been matched. Refer to our earlier paper~\cite{peg-practical} for more details.

Interestingly, this strategy can be implemented entirely within the stateful
parsing paradigm. In particular, when we speak of \emph{result of a parse}, we
are really referring to a \emph{delta} of the parse state. These deltas need to
be stored in the parse state, so that they can be retrieved by left-recursive
invocations. We also need to track, within the state, which left-recursive
parsers have been invoked at which input position, so that we may recognize
left-recursive invocations.\footnote{We could track the whole state instead of
  the input position, but we enforce the stronger requirement that a parser
  invocation has to advance the input position not to be considered
  left-recursive: unlike other state changes, an increase in input position is a
  strong proof of progress.} 

Recognizing left-recursive invocations is the task of a dedicated parser that
must be wrapped around all left-recursive parsers. We do so by annotating
recursive parsers with the \texttt{!} operator. Since left-recursion requires
recursive references (cf. Section~\ref{autumn-api}), we can easily check that
all such parsers have been properly annotated: a missing annotation will result
in an unresolved reference instead of a mystifying infinite loop.

%%%%%%%%%%%%%%%%%%%%%%%%%%%%%%%%%%%%%%%%%%%%%%%%%%%%%%%%%%%%%%%%%%%%%%%%%%%%%%%%
\section{Use Case}
\label{usecase}

In this section, we demonstrate the stateful parsing approach with a realistic
use case, using our Autumn library. We implement a parser for a simple, yet
non-trivial, statically-typed, object-oriented programming language, which we
call \emph{Examply}. This imaginary language draws inspiration from Java (its
main unit of definition is the class), Kotlin (its postfix type notation and
closure notation) and Python (significant indentation). Its full grammar,
written with Autumn, can be found online~\cite{sle2016artif}. Examply possesses
two common context-sensitive features:

\begin{itemize}

\item \textbf{Significant whitespace} --- Indentation is significant: Like
  Python, the language does not use curly braces or keywords to delimit blocks
  of code such as loop or function bodies. Instead, these constructs expect to
  have their body indented with respect to their first line. Similarly, a
  decrease in indentation signifies the end of the block. Newlines are also
  significant, as they are used to separate successive statements and
  declarations.

\item \textbf{Namespace classification} --- The parser needs to know which
  identifiers refer to types. In our language, this is needed because there is
  an ambiguity between the syntax of function calls --- which can receive a
  closure parameter as an indented block --- and the syntax of anonymous
  classes:

\begin{lstlisting}
val a = myFunction()
    myFunction2()

val b = MyClass()
    var x: Int
    fun foo() { ... }
\end{lstlisting}

  The body of a class only admits declarations, while the block part of a
  function call admits all statements (including declarations). These two
  constructs result in different nodes being added to the abstract syntax tree
  (AST) produced by the parse.

\end{itemize}

%%%%%%%%%%%%%%%%%%%%%%%%%%%%%%%%%%%%%%%%
\subsection{Significant Whitespace}
\label{whitespace}

We now explain how Examply handles significant whitespace. The code enabling
this feature is shown in Figure~\ref{indent-code}.

The usual whitespace handling strategy is to assume that every parser consumes
its trailing whitespace through invocation of the \texttt{whitespace} parser. To
do so, it is only necessary to ensure that all ``primitive'' parsers
(fullfilling what is traditionally the role of lexical analysis: matching
identifiers, literals, keywords and operators) consume their trailing
whitespace: then all parsers will do so by transitivity. The \texttt{Grammar}
class provides some support for this, including lexical analysis emulation (not
demonstrated in this paper), and the [\texttt{+"lit"}] syntax which evaluates to
a parser matching a literal string and any trailing whitespace. Ultimately, this
leads to an important guarantee: all parsers are invoked at an input position
where no leading whitespace is present.

In order to handle significant whitespace, we maintain two data structures. The
first one, \texttt{IndentMap}, maps line numbers to two quantities: the input
position at which the indentation ends on that line, and the indentation count,
which is obtained by expanding tabs to tab stops aligned to multiples of 4. The
second data structure, \texttt{IndentStack}, is a stack that stores the
indentation counts for all enclosing blocks.

\texttt{Context.indent} and \texttt{Context.istack} are extension properties for
the \texttt{Context} class and provide syntactic sugar to access the indentation
count on the current line, and the indentation stack, respectively.

We build \texttt{IndentMap} at the beginning of the parse, through the
invocation of the \texttt{buildIndentMap} parser. This does not require any
special tricks: each parser has access to the whole input (\texttt{ctx.text}).
The parser does not advance the input position (\texttt{ctx.pos}), so that
subsequent parsers are free to proceed.

Within the grammar, we use the \texttt{indent} parser to require an indented
block, and the \texttt{dedent} parser to test for the end of an indented block.
The implementation of these parsers is straightforward. \texttt{indent} checks
if the current line is indented with respect to the indentation of the current
block (the top of \texttt{IndentStack}, initialized to 0). If so, it succeeds
after pushing the new indentation count onto the stack. \texttt{dedent} checks
to see if the indentation count of the current line is less than that of the
current block, or if we have reached the end of the input. If so it succeeds,
after popping the previous count from the stack.

Finally, the \texttt{newline} parser succeeds if and only if it is invoked at
the end of the indentation on the current line or at the end of the input.

\begin{figure}[!ht]
\begin{lstlisting}
data class IndentEntry (val count: Int, val end: Int)

class IndentMap: InertState<IndentMap> {
  lateinit var map: Map<Int, IndentEntry>
  fun get(ctx: Context): IndentEntry =
    map[ctx.lineMap.lineFromOffset(ctx.pos)]!!
}

class IndentStack: StackState<Int>()

val Context.indent: IndentEntry
  get() = state(IndentMap::class).get(this)
val Context.istack: IndentStack
  get() = state(IndentStack::class)

val buildIndentMap = Parser { ctx ->
  val map = HashMap<Int, IndentEntry>()
  var pos = 0
  ctx.text.split('\n').forEachIndexed { i, str ->
    val wspace = str.takeWhile {
      it == ' ' || it == '\t' }
    val count = wspace.wspace.expandTabs(4).length
    map.put(i, IndentEntry(count, pos + wspace.length))
    pos += str.length + 1
  }
  ctx.state(IndentMap::class).map = map
  Success
}

val indent = Parser { ctx ->
  val new = ctx.indent.count
  val old = ctx.istack.peek() ?: 0
  if (new > old) Success after { ctx.istack.push(new) }
  else ctx.failure {
    "Expecting indentation > $old positions" }
}

val dedent = Parser { ctx ->
  val new = ctx.indent.count
  val old = ctx.istack.peek() ?: 0
  if (new < old || ctx.pos == ctx.text.length - 1)
    Success after { ctx.istack.pop() }
  else ctx.failure {
    "Expecting indentation < $old positions" }
}

val newline = Predicate {
  indent.end == pos || ctx.pos == ctx.text.length - 1 }
\end{lstlisting}
  \caption{Using Autumn to implement significant whitespace handling for
    Examply.}
\label{indent-code}
\end{figure}

%%%%%%%%%%%%%%%%%%%%%%%%%%%%%%%%%%%%%%%%
\subsection{Namespace Classification}
\label{nstrack}

Examply needs to distinguish between types (i.e., class names) and other
identifiers at parse-time, in order to resolve ambiguities. We call this process
\emph{namespace classification}: we want to know whether an identifier belongs
to the namespace of types or not. Such parse-time tracking is reminiscent of the
C language, and is seldom seen in modern languages. In Examply, we could avoid
it by adding a keyword (e.g., \texttt{new}) to disambiguate constructor
invocation from function invocation. Or we could perform AST-disambiguation
passes after parsing. We stress that Examply is designed to showcase the
strength of the principled stateful approach, and notably its ability to deal
with the quirks of existing languages.

To understand the logistics of namespace classification, we must first define
how our imaginary language handles type references:

\begin{enumerate}

\item Types are always referenced through a single identifier, except within
  imports where they are preceded by a package string.

\item A class name can only be referenced after or within its definition.

\item A class definition can appear anywhere other declarations can: at the
  top-level, within another class, or within a code block.

\item Class definitions are lexically scoped: a class has access to all imported
  classes and to all classes defined before it within one of its outer scopes
  (class body or code block).

\item A class has access to all classes defined within its superclass and other
  ancestors.

\item A class cannot inherit from one of its outer classes.

\item In order to avoid ambiguous type names (for instance, both a class defined
  in an outer class and a class defined in a superclass could bear the same
  name), Examply features type aliases that assign an alternate name to an
  existing type. Type aliases can appear anywhere a class definition can appear,
  and have the same visibility as class definitions.

\end{enumerate}

Figure~\ref{type-code} shows the code handling namespace classification. We do
not engage with its minutiae, but instead give a high-level description of its
operation. The code itself should demonstrate that the implementation of these
ideas is terse and readable, even to those who ignore the precise semantics of
some operations. We note however that in the code, \texttt{ctx.stack} refers to
the stack used to construct AST nodes. We occasionally peek in this stack in
order to retrieve identifiers.

The main data structure is the \texttt{TypeStack} \texttt{State} instance. It
holds a stack of \texttt{Type}s, which are pairs formed by a string and a list
of other \texttt{Type}s. Intuitively, each \texttt{Type} instance represents a
class name alongside with a list of classes accessible through it: its inner
classes and the inner classes of its ancestors. We call these classes the
\emph{private classes} of a class: they cease being accessible once the class
definition ends. We define two helper functions over the type stack:
\texttt{isType} checks if an identifier refers to a type, and \texttt{priv}
returns the private classes of the named class, or an empty list if no such
class exists.

Each time we encounter a new type during the parse, it is pushed onto the type
stack. This is the the task of the \texttt{NewType} parser, which is applied to
identifiers introduced by classes and type aliases. A parameter controls whether
the new type is an alias, in which case it inherits the private classes of the
aliased class. Note that classes start with an empty list of private classes.
This will be updated once the class definition is complete.

Once a scope (a class body, or some code block) is exited, the types introduced
within it are not longer accessible. To enforce this, we use the \texttt{Scoped}
parser: it saves the type stack size, invokes its child parser (corresponding to
a scope), then removes any extraneous items from the type stack.

If the scoped body was a class body, the \texttt{Type} representing the class on
the stack must be updated with a list of its private classes, so that an inner
class may access them. This is the role of the \texttt{ClassDef} parser. It
looks up the class and superclass names on the AST stack, then looks up the list
of private classes of the superclass. If found, this list is pushed on the type
stack. All of this is done after taking a snapshot of type stack. Subsequently,
the \texttt{body} parser is invoked and, if successful, a delta of the type
stack is generated using the snapshot. This delta corresponds to the private
classes of the class, including those introduced by its superclass. The snapshot
is restored and the topmost entry on the type stack (which represents the class)
is removed and replaced with a new one that binds the class name to its private
classes.

A reduced version of this process also needs to happen for anonymous classes:
they need to access their superclass' private classes, but no \texttt{Type}
record must be created for them. The \texttt{anonClassInherit} parser fullfills
this role, by reusing the \texttt{inherit} function.

Finally, to resolve the ambiguity, we use the \texttt{classGuard} parser: it
performs a lookahead, attempting to match an identifier, and succeeding only
this identifier refers to a type.

We defer our assessment of the approach until Section~\ref{discussion}.

\begin{figure}
\begin{lstlisting}
data class Type (val name: String, val priv: LinkList<Type>)
class TypeStack: MonotonicStack<Type>()
val Context.types: TypeStack
    get() = state(TypeStack::class)

fun isType(ctx: Context, iden: String): Boolean
  = ctx.types.stream().any { it.name == iden }

fun priv(ctx: Context, iden: String): LinkList<Type>
  = ctx.types.stream()
    .filter { it.name == iden }
    .next() ?.priv ?: LinkList()

fun NewType (child: Parser, alias: Boolean = false)
= Parser { ctx ->
    child.parse(ctx) andDo {
      val name = ctx.stack.peek() as String
      val priv = if (alias) priv(ctx, name)
                 else LinkList()
      ctx.types.push(Type(name, priv))
  } }

fun Scoped(body: Parser) = Parser { ctx ->
  val size = ctx.types.size
  body.parse(ctx) andDo { ctx.types.truncate(size) }
}

fun inherit(ctx: Context, name: String)
  = priv(ctx, name).stream().each { ctx.types.push(it) }

fun ClassDef (body: Parser)
= Parser { ctx ->
    val parent = ctx.stack.at(0) as Maybe<SimpleType>
    val name = ctx.stack.at(1) as String
    val snapshot = ctx.types.snapshot()
    if (parent is Some<SimpleType>)
      inherit(ctx, parent.value.name)
    body.parse(ctx) andDo {
      val diff = ctx.types.diff(snapshot)
      ctx.types.restore(snapshot)
      ctx.types.pop()
      ctx.types.push(Type(name, diff))
} }

val anonClassInherit = Perform { ctx ->
  inherit(ctx, ctx.stack.at(1) as String) }

val classGuard = Seq(iden, Predicate { ctx ->
  isType(ctx, ctx.stack.peek() as String) }
).ahead
\end{lstlisting}
  \caption{Using Autumn to implement namespace classification for Examply.}
\label{type-code}
\end{figure}

%%%%%%%%%%%%%%%%%%%%%%%%%%%%%%%%%%%%%%%%
\subsection{Putting it all together}
\label{together}

To illustrate the use of significant whitespace and namespace classification (as
presented in sections \ref{whitespace} and \ref{nstrack}), let's look at two
short examples. First, here is how we define an indented code block:

\begin{minipage}{\textwidth}
\begin{lstlisting}
val statements = 
  Seq(indent, Scoped(!"statement" until dedent))
      .collect<Stmt>()
\end{lstlisting}
\end{minipage}

The block starts by an increase in indentation (\texttt{indent}), and ends when
a decrease in indentation is encountered (\texttt{dedent}), parsing statements
in the mean time (\texttt{!"statement" until dedent}). The \texttt{collect} part
instructs the parser to collect all statement nodes pushed onto the AST stack
and to aggregate them in a list, which is itself pushed onto that stack. We also
see that all indented statements form a scope (\texttt{Scoped}) in the sense of
Section~\ref{nstrack}.

Second, here is how the parser for ``block constructors'' (i.e., anonymous
classes) is defined:

\begin{lstlisting}
val blockCtorBody = Scoped(Seq(anonClassInherit, decls))
val blockCtor
  = Seq(classGuard, iden, paramList, blockCtorBody)
    .build { CtorCall(get(), get(), get()) }
\end{lstlisting}

The parser is simply guarded with the \texttt{classGuard} parser, which checks
if there is an identifier at the current input position, and whether this
identifier refers to a type. The body of the class can access the superclass'
private classes through \texttt{anonClassInherit}. The \texttt{Scope} wrapper
ensures that this access is restricted to the class and does not spread to the
code that follows. \texttt{decls} refers to an indented block of declarations.

%%%%%%%%%%%%%%%%%%%%%%%%%%%%%%%%%%%%%%%%
\subsection{Discussion}
\label{discussion}

We have implemented two context-sensitive features in for an imaginary but
non-trivial programming language. With the code we have shown so far, the rest
of the grammar is trivially able to define code delimited by changes in
indentation, or by newlines (Section~\ref{whitespace}); or to direct the parse
depending on whether an identifier refers to a type (Section~\ref{nstrack}).

All this, by itself, is no mean feat. There are few parsing tools where this is
possible to begin with (most of them are presented in Section~\ref{related}).
Significant indentation handling, in particular, is non-trivial because Autumn
does not include a lexical analysis layer out of the box.

We also underline that the presented implementations are rather terse, less than
50 lines of code each. Some of that does come from our choice of implementation
language, but it also shows that the principled stateful parsing approach does
not impose a significant boilerplate overhead. In particular, the ability to
reuse state-handling strategies, such as \texttt{MonotonicStack} means that
\emph{context-transparency} comes almost for free.

The state manipulations operations from sections~\ref{intuition}
and~\ref{operations} are strangely discreet in our examples. Significant
whitespace handling does not use them at all, while namespace classification
performs a \texttt{diff} in order to capture the types introduced by a class
body. But because they do not appear in the code does not mean the operations
are not used, they are simply hidden from view. The basic contract of principled
stateful parsing is that each parser either succeeds or leaves the state
untouched. You can convince yourself that all the parsers and parser combinators
we introduced satisfy this condition, either by reusing existing combinators, or
by delegating the responsibility for this to their single subparser.

It remains that the parsers we introduce do get backtracked over during the
parse. As such, their state mostly get saved and restored by other parsers that
invoke them, directly or indirectly. It is in fact crucial for the state to get
restored when backtracking occurs: we need to know the correct indentation level
whenever we backtrack out of a block; we also need to know which identifiers are
classes, even when backtracking over a type definition. Granted, given that most
constructs in the grammar are guarded by specific keywords, such backtracking
occurences should be rare. However, unlike the other, often \emph{accidentally}
stateful parsers (cf. Section~\ref{related}), context transparency ensures that
we can evolve the grammar as we see fit, without fear of breaking the mechanisms
we just introduced. It is also a pre-requisite for safe grammar composition.

In fact, the scarcity of state operations is a boon: it means that the benefits
of our approach come at very little cost, at least implementation-wise. We would
also disabuse the reader of the notion that the Autumn codebase hides some
devilish complexity in order to make up for this: the whole
library~\cite{sle2016artif} is less than 2500 lines of code. All pre-defined
parsers live in a single file of less than 500 lines. This file defines around
50 parsers: those corresponding to all basic PEG~\cite{peg} operators, as well
as numerous extensions, notably to work with error messages, AST nodes, ...

%%%%%%%%%%%%%%%%%%%%%%%%%%%%%%%%%%%%%%%%
\subsection{Performance}

Performance has not been our focal point, but preliminary testing seems to
indicate that performance is within an order of magnitude of mainstream parsing
tools such as Rats!~\cite{rats} and Parboiled~\cite{parboiled} for context-free
grammars. The implementation currently incurs overhead even for context-free
grammars, which we are working to reduce. The overhead scales with the amount of
state in use, depending on the \texttt{State} implementation details. The costly
operations are the creation of snapshots and deltas. In general, memory
allocations tend to be the bottleneck, so increased sharing between snapshots
results in better performances. Indeed, we've had success with purely functional
data structures~\cite{okasaki}.

Just like PEG parsing without full memoization, parsing has exponential
complexity in the worst case. In practice however run times are acceptable, as
programming language grammars are fairly deterministic.

%%%%%%%%%%%%%%%%%%%%%%%%%%%%%%%%%%%%%%%%%%%%%%%%%%%%%%%%%%%%%%%%%%%%%%%%%%%%%%%%
\section{Conclusion}
\label{conclusion}

In this paper, we proposed an approach to tackle the problem of
context-sensitive parsing. Our solution, unlike existing ones, possesses the
property of \emph{context transparency}: grammatical constructs are unaware of
the context shared between their ancestors and their descendants, making it
easier to write, evolve and compose context-sensitive grammars.

We proceeded in two parts. First, we allowed parsers to manipulate a mutable
data store, so as to enable context-sensitivity through \emph{recall}. Second,
we required parsers to behave transactionally: a parser must either succeed, or
fail and leave the state unaltered. This transactional discipline, which we call
\emph{principled stateful parsing}, prevents parsing mechanisms such as
backtracking and memoization to break the guarantee of context transparency.

To enforce the principled stateful parsing discipline, we supplied formally
specified state manipulation operations. The role of these operations is to
snapshot and restore the state, as well as to create and merge deltas between a
snapshot and the current state.

We implemented our approach in a parsing library called \emph{Autumn}, and
showed how it can be used in practice to implement common context-sensitive
grammar features such as significant whitespace and namespace classification. We
underline the low boilerplate and conceptual overhead introduced by the
approach.

We view this work as a first step towards bringing disciplined grammarware
engineering (as defined by Klint, Lämmel and Verhoef~\cite{grammarware}) to
context-sensitive parsing.

\newpage

%%%%%%%%%%%%%%%%%%%%%%%%%%%%%%%%%%%%%%%%%%%%%%%%%%%%%%%%%%%%%%%%%%%%%%%%%%%%%%%%

% \acks
% Acknowledgments, if needed.

%%%%%%%%%%%%%%%%%%%%%%%%%%%%%%%%%%%%%%%%%%%%%%%%%%%%%%%%%%%%%%%%%%%%%%%%%%%%%%%%

%%%%%%%%%%%%%%%%%%%%%%%%%%%%%%%%%%%%%%%%%%%%%%%%%%%%%%%%%%%%%%%%%%%%%%%%%%%%%%%%

\end{document}